%% file: Main_Quantum.tex
\documentclass[journal]{IEEEtran}
\IEEEoverridecommandlockouts
\usepackage{cite}
\usepackage{multirow} 
\usepackage{amsmath,amssymb,amsfonts}
\usepackage{lipsum,multicol}
\usepackage{braket}
\usepackage{comment}

\usepackage{algorithmic}
\usepackage{graphicx}
\usepackage{algorithmic}
\usepackage[ruled,vlined]{algorithm2e}
\SetKwInput{KwInput}{Input}  

\SetKwInput{KwOutput}{Output}              
\SetKw{Break}{break}
\SetKwRepeat{Do}{do}{while}
\usepackage{textcomp}
\usepackage{float}
\usepackage{xcolor}
\usepackage{stfloats}
\usepackage{empheq}

\def\BibTeX{{\rm B\kern-.05em{\sc i\kern-.025em b}\kern-.08em
    T\kern-.1667em\lower.7ex\hbox{E}\kern-.125emX}}

\usepackage[acronym]{glossaries}
\makeglossaries
\input{Acronyms}
\begin{document}

\title{Hybrid Quantum–Classical Detection for RIS-Assisted SC-FDE via Grover Adaptive Search\\ \vspace{0.5cm} \normalsize Maryam Tariq, Omar Alhussein, Raneem Abdelraheem, Abdullah Quran, Georges Kaddoum, and Sami Muhaidat   \vspace{-1.0cm}}
\author{

\thanks{ M. Tariq, O. Alhussein, R. Abdelraheem, A. Quran, and S. Muhaidat are with the 6G
Research Center, College of Computing and Mathematical Sciences, Khalifa University, Abu Dhabi, UAE. (email:\{100036622, omar.alhussein, 100067007, 100062547, sami.muhaidat\}@ku.ac.ae).}
\thanks{G. Kaddoum is with the Department of Electrical Engineering, 
\'Ecole de Technologie Sup\'erieure (\'ETS), Universit\'e du Qu\'ebec, 
Montr\'eal, QC H3C 1K3, Canada. (email: georges.kaddoum@etsmtl.ca).}
}






\maketitle

\bstctlcite{IEEEexample:BSTcontrol}

\begin{abstract}
Wideband and low-latency requirements in sixth-generation (6G) networks demand detectors that approach maximum-likelihood (ML) performance without incurring exponential complexity. This work develops a hybrid quantum–classical detection framework for reconfigurable intelligent surface (RIS)-assisted single-carrier (SC) frequency-domain equalization (FDE) over frequency-selective channels. The ML detection objective is reformulated as a quadratic unconstrained binary optimization (QUBO) problem and solved via Grover adaptive search (GAS). To accelerate convergence, we introduce a frequency-domain MMSE threshold that exploits the circulant structure of SC-FDE channels, yielding low-complexity initialization. The framework is evaluated across varying channel lengths and RIS sizes, confirming robustness and scalability. In addition, GAS requirements are quantified through register widths and gate counts, and its query complexity is analyzed to characterize the algorithm’s cost for block transmission in frequency-selective channels. Quantum circuit simulations are conducted in Qiskit under both ideal and noisy conditions. In the ideal case, the detector achieves near-optimal performance while benefiting from Grover’s quadratic speedup, reducing the search cost from $\mathcal{O}(M^{N})$ exhaustive evaluations to $\mathcal{O}(\sqrt{M^{N}})$ oracle queries. Under noise, the shallow depth of the GAS circuits, aided by MMSE initialization, makes depolarizing errors negligible, while readout errors introduce moderate degradation yet still preserve performance close to the MMSE baseline. These results establish the feasibility of quantum-enhanced detection for RIS-assisted broadband communications, highlighting both algorithmic scalability and practical robustness for 6G networks.
\end{abstract}

\vspace{0.25em} 


\section{Introduction}
Envisioned \gls{6G} wireless networks will enable emerging applications such as real-time remote surgery, connected autonomous vehicles, and ultra-high-speed wireless backhaul~\cite{kharche20236G, he20206g, jiang2024terahertz}. These use cases impose stringent physical-layer requirements, notably in terms of \gls{URLLC} and multi-gigabit data throughput. To satisfy these demands, future systems are expected to exploit wide bandwidths at high-frequency spectrum, including \gls{mmWave} and sub-Terahertz (sub-THz) bands~\cite{chukhno2023models}.

However, operation in such regimes presents fundamental physical-layer challenges. First, wideband transmission induces \gls{ISI} due to the frequency-selective nature of broadband fading channels. In ultra-high data rate scenarios ranging from $100$~Gbps to $1$~Tbps, even modest microsecond-scale delay spreads can lead to \gls{ISI} spanning thousands of symbols~\cite{tusha2024interference}. Second, high-frequency signals experience challenging propagation conditions, making reliable reception difficult under \gls{NLOS} scenarios~\cite{amodu2024technical}. These effects collectively degrade detection performance and increase the receiver's computational burden. Consequently, robust transceiver architectures are required to jointly mitigate temporal dispersion and propagation impairments, while maintaining scalability in complexity.

A promising approach is the integration of \gls{SC}-\gls{FDE} with \gls{RIS}~\cite{pancaldi2008single,9140329}, offering an efficient and reliable physical-layer solution for wideband frequency-selective channels~\cite{tariq2025ris, li2023reconfigurable,nagaya2024reduced,mainref}. \gls{SC}-\gls{FDE} employs block-based transmission, appending a \gls{CP} to each block to convert linear convolution into circular convolution, thereby enabling low-complexity equalization via the \gls{DFT}~\cite{4155121}. This approach significantly reduces receiver complexity compared to conventional time-domain equalization (TDE) while retaining strong resilience to ISI~\cite{falconer2002frequency}. Moreover, the inherently low \gls{PAPR} and reduced sensitivity to \gls{CFO} make \gls{SC}-\gls{FDE} particularly well-suited for high-frequency, high-throughput scenarios, where power-amplifier efficiency and synchronization robustness are critical~\cite{falconer2002frequency,pancaldi2008single}.

On the other hand, \gls{RIS} enables programmable control of the wireless propagation environment by dynamically adjusting the phase shifts of large arrays of passive, sub-wavelength elements, thereby improving received signal strength, extending coverage, and enhancing link robustness under challenging conditions~\cite{Jung2021On,Elmossallamy2020Reconfigurable}. In frequency-selective channels, \gls{RIS} can be configured to align reflected signal phases with dominant channel taps~\cite{tariq2025ris} or to introduce virtual propagation paths that enhance multipath diversity~\cite{mainref}, thus complementing \gls{SC}-\gls{FDE} by shaping the delay-domain energy distribution. However, in wideband scenarios, larger delay spreads necessitate longer \gls{CP}s to prevent inter-block interference (IBI) and preserve the channel's circulant structure. Since both the CP and the block length $N$ scale with the channel memory length, highly frequency-selective channels directly lead to larger $N$, causing the complexity of optimal maximum-likelihood detection to grow exponentially.

Conventional detection algorithms for frequency-selective channels involve trade-offs between computational complexity and detection accuracy. \Gls{MLD} achieves optimal performance but requires an exhaustive search over all possible symbol combinations, resulting in exponential complexity on the order of $\mathcal{O}(M^N)$, where $M$ is the constellation size~\cite{1269977}. For finite-memory channels, the Viterbi algorithm offers an efficient implementation of MLD by exploiting the channel's Markovian structure via trellis-based decoding, reducing the complexity to $\mathcal{O}(N M^{L-1})$, where $L$ denotes the number of channel taps~\cite{viterbi_comp}. However, the Viterbi algorithm's complexity still scales exponentially with $L$ and $M$, rendering it impractical for highly dispersive or high-order modulation systems. To further reduce complexity, sphere decoding (SD) has been proposed as a near-optimal alternative \cite{1408197}. While it reduces the average-case complexity relative to MLD, its runtime is data-dependent and can still be prohibitive under low-SNR or long-delay spread conditions. Fixed-complexity variants such as the fixed-complexity sphere decoder (FSD) and K-best algorithms impose a structured search strategy to bound complexity, but often suffer from performance loss and require heuristic tuning~\cite{5378999}. Linear detectors such as zero-forcing (ZF) and minimum mean square error (MMSE) offer lower complexity, typically $\mathcal{O}(N \log N)$ when implemented using \gls{DFT}-based equalization, but are prone to noise amplification and residual \gls{ISI} in severely frequency-selective channels~\cite{6259859}. Despite the diversity of approaches, existing methods either compromise optimality or exhibit computational demands that limit scalability. These limitations motivate the development of advanced detection techniques, such as quantum-enhanced methods, that can simultaneously offer near-optimal performance and low predictable complexity under practical channel conditions.

Quantum computing offers a fundamentally new paradigm for addressing the combinatorial complexity of signal detection in next-generation wireless systems~\cite{10032075,9915359}. A key tool in this domain is Grover's search algorithm~\cite{grover1996fast}, which achieves a quadratic speedup over classical exhaustive search by amplifying the probability amplitude of valid solutions through repeated queries to a quantum oracle. In the context of \gls{MLD}, Grover's algorithm reduces the search complexity from $\mathcal{O}\!\left(M^{N}\right)$ for exhaustive classical evaluation to $\mathcal{O}\!\left(\sqrt{M^{N}}\right)$ in terms of quantum oracle calls.
 To extend Grover's framework to structured binary optimization, Gilliam \emph{et al.} proposed the Grover adaptive search (GAS) algorithm~\cite{bulger2003implementing}, which guides the amplitude amplification process using a cost-function oracle tailored to the problem structure. GAS formulates such problems as instances of quadratic unconstrained binary optimization (QUBO), enabling execution on Ising-based quantum optimizers. To further improve scalability and reduce quantum resource overhead, Gilliam \emph{et al.} introduced a dictionary-based encoding scheme for efficiently representing polynomial cost functions with integer coefficients in quantum circuits~\cite{gilliam2020optimizing}. This construction simplifies oracle implementation and broadens the applicability of GAS to a wide range of combinatorial problems, including symbol detection in wireless communication systems.

Although GAS was initially designed for fault-tolerant quantum computing (FTQC), recent studies have explored its adaptation to noisy intermediate-scale quantum (NISQ) devices in small-scale scenarios~\cite{Wang2020}. However, its performance in realistic noisy settings remains an open question, particularly for detection problems in wireless systems. Grover-based algorithms such as GAS are inherently sensitive to noise accumulation across iterative amplitude amplification steps, and quantum noise, including decoherence, gate infidelity, and measurement error~\cite{mcclean2021limitations,10200430}. In NISQ hardware, these imperfections can distort oracle queries and hinder convergence, especially in circuits with many controlled operations~\cite{QuantumInNISQ}. GAS circuits are especially vulnerable in this regard, as both the oracle and diffusion operators involve multiple layers of controlled gates.

To improve GAS robustness to quantum noise on NISQ hardware, circuit depth must be constrained and noise exposure minimized. Adaptive initialization strategies, such as MMSE-based thresholding~\cite{6730885}, serve this purpose by setting an initial threshold close to the optimal value, thereby reducing the required number of iterations and resulting in shallower circuits. Nevertheless, the sensitivity of GAS to specific noise models is not yet thoroughly investigated. This motivates a comprehensive investigation of hybrid quantum–classical detection under realistic noise conditions, focusing on moderate-depth circuits for practical search spaces tractable on current simulators and hardware.

Several studies have explored the application of Grover-based quantum algorithms to wireless signal detection, aiming to accelerate MLD in unstructured search problems. Botsinis~\emph{et al.} first demonstrated that MLD could be significantly sped up using Grover's algorithm on quantum hardware~\cite{botsinis2013quantum}. This led to the adoption of Grover-inspired techniques such as the Boyer–Brassard–Høyer–Tapp (BBHT)~\cite{boyer1998tight} and Dürr–Høyer (DH)~\cite{durr1996quantum} algorithms, which were further tailored to multiuser scenarios via MMSE-initialized thresholds to reduce the expected number of oracle queries~\cite{6730885}.

More recent work has focused on adapting GAS to structured detection problems. Norimoto~\emph{et al.} formulated MIMO MLD as a higher-order unconstrained binary optimization (HUBO) problem and proposed a GAS-based solution with efficient circuit-level implementation~\cite{10044091}. Subsequent studies extended this approach to power-domain non-orthogonal multiple access (NOMA) systems with adaptive thresholding~\cite{10200430}, and to generalized spatial modulation (GSM), where MLD was reformulated as a polynomial optimization problem and solved using GAS under the assumption of FTQC~\cite{GASonSM}. 

While previous studies \cite{botsinis2013quantum, boyer1998tight, durr1996quantum,6730885,10044091,10200430,GASonSM} have demonstrated the potential of quantum-assisted MLD in wireless systems, they have focused mainly on flat-fading channels. Many of these studies also assume the availability of FTQC, neglecting the limitations imposed by quantum noise in near-term devices. To date, the specific challenges posed by frequency-selective propagation, particularly in RIS-assisted environments, remain unexplored. This work presents, to the best of our knowledge, the first quantum-assisted detector explicitly designed for RIS-assisted broadband systems, comparing its performance in both ideal and noisy regimes. The proposed approach formulates a real-valued QUBO problem that can be explicitly represented by a quantum circuit and employs a low-complexity frequency-domain MMSE estimator as the initial threshold, in contrast to existing literature that relies on spatial-domain MMSE. It should be noted that a preliminary version of this work appeared in \cite{tariq2025hybrid}, where initial results were presented. This paper extends the previous work by providing a detailed explanation of the quantum circuit construction, quantifying the required resources, presenting additional performance results, and evaluating the system under both ideal and noisy regimes. The main contributions of this paper are summarized as follows:

\begin{itemize}
    \item This work develops a hybrid quantum–classical detection framework for \gls{RIS}-assisted \gls{SC}-\gls{FDE} systems over broadband frequency-selective \gls{NLOS} channels. The \gls{BPSK} \gls{MLD} objective function is expressed in the frequency domain and formulated as a QUBO problem, which is mapped to a quantum circuit with explicit register widths and gate counts, allowing for direct solution via GAS.
    
    \item The proposed detector is implemented using a quantum circuit comprising state preparation, controlled-phase rotation cost-function oracle, and diffusion operator. A frequency-domain MMSE-based thresholding strategy is introduced for low-complexity initialization, leveraging \gls{SC}-\gls{FDE} equalization to reduce the quantum search space and circuit depth, thereby accelerating convergence and improving noise robustness.
    
    \item The framework is evaluated using Qiskit quantum-circuit simulations under both ideal and noisy regimes, including depolarizing and readout errors, to assess performance and feasibility under realistic quantum-hardware scenarios, which have been overlooked in conventional studies that assume the realization of FTQC.
\end{itemize}

The remainder of this paper is organized as follows. Essential quantum computing preliminaries are reviewed in Section~\ref{sec:Background}. The system model for RIS-assisted SC-FDE transmission, including transmitter design, channel model, and receiver architecture, is described in Section~\ref{sec:system_model}. Section~\ref{Implementation} formulates the MLD problem as a QUBO and outlines its solution via Grover search. GAS-based detection is detailed in Section~\ref{sec:GAS}, followed by a low-complexity MMSE-based preprocessing step for threshold initialization in Section~\ref{sec:threshold}. Section~\ref{sec:Circuit} describes the corresponding GAS circuit implementation, while section~\ref{sec:resource} then presents a resource analysis, quantifying GAS query complexity, register widths and gate counts.
 The considered quantum noise models and their impact on hybrid detection are discussed in Section~\ref{sec:quantum_noise}, while Section~\ref{Results} reports simulation results under both ideal and noisy conditions. The paper concludes with Section~\ref{conc}.

\section{Quantum Preliminaries}
\label{sec:Background}
This section summarizes the core quantum concepts underpinning the proposed hybrid detection scheme, including qubits and superposition,  quantum gates, and Grover's search algorithm. Emphasis is placed on elements directly relevant to the detection framework presented in this work.
\subsection{Qubits and Superposition}
A qubit is a two-level quantum system represented by a linear combination of computational basis states:
\begin{equation}
\ket{\psi} = \alpha\ket{0} + \beta\ket{1}, \quad \text{with } |\alpha|^2 + |\beta|^2 = 1,
\end{equation}
where $\alpha, \beta \in \mathbb{C}$. Multi-qubit systems are modeled by tensor products of individual qubits, enabling quantum parallelism via superposition. Upon measurement, the system probabilistically collapses to one of the basis states.
\subsection{Quantum Gates and Circuits}
Quantum operations are represented by unitary transformations acting on qubit registers. In this work, the primary gates used are:
\begin{itemize}
    \item {Hadamard ($H$):} Creates equal superposition, e.g., $H\ket{0} = (\ket{0} + \ket{1})/\sqrt{2}$.
    \item {Pauli gates ($X$, $Y$, $Z$):} Perform bit and/or phase flips on a single qubit.
    \item {Controlled-NOT (CNOT):} A two-qubit entangling gate that applies $X$ on the target qubit conditional on the control qubit being $\ket{1}$.
    \item {Phase rotation $R_z(\theta)$:} Applies a rotation about the $z$-axis of the Bloch sphere,
\begin{equation}
R_z(\theta) =
\begin{bmatrix}
1 & 0 \\
0 & e^{i\theta}
\end{bmatrix}.
\end{equation}
Controlled versions of $R_z(\theta)$ are denoted as CR (single-controlled) or CCR (double-controlled), with higher-order multi-controlled rotations also possible. These are used to encode linear and quadratic terms in the QUBO formulation.
\end{itemize}
\subsection{Grover's Search Algorithm}
Grover's search enables the retrieval of marked states from an unstructured search space of $S$ unsorted elements using $\mathcal{O}(\sqrt{S})$ oracle queries.
 The algorithm starts by preparing a uniform superposition over all basis states as follows
\begin{equation}
\ket{\psi_0} = H^{\otimes n} \ket{0}^{\otimes n}.
\end{equation}
Each Grover iteration, $G = D\,O$, consists of:
\begin{itemize}
    \item {Oracle ($O$):} Applies a phase flip to the marked states. 
    \item {Diffusion operator ($D$):} Amplifies the probability of the marked states by inverting all amplitudes about their mean.
\end{itemize}
In this work, the search space has size $S=M^{N}$ with $n=N\log_{2}M$ qubits.
The phase oracle encodes the QUBO form of the detection metric, so Grover iterations amplify low-cost symbol vectors. The solution density and the size of the search space determine the number of iterations. 
The optimal number of Grover iterations that maximizes the success probability is given by~\cite{brassard2000quantum}
\begin{equation}
L_{\mathrm{opt}} = 
\left\lfloor 
\frac{\pi}{4} \sqrt{\frac{S}{M_s}} 
\right\rfloor ,
\end{equation}
where $M_s$ is the number of marked solution states. Hence, with a unique solution, $M_s=1$, the algorithm requires $\mathcal{O}\!\left(\sqrt{M^{N}}\right)$ oracle calls.
\section{System Model}
\label{sec:system_model}
\begin{figure}
    \includegraphics[width=1\linewidth]{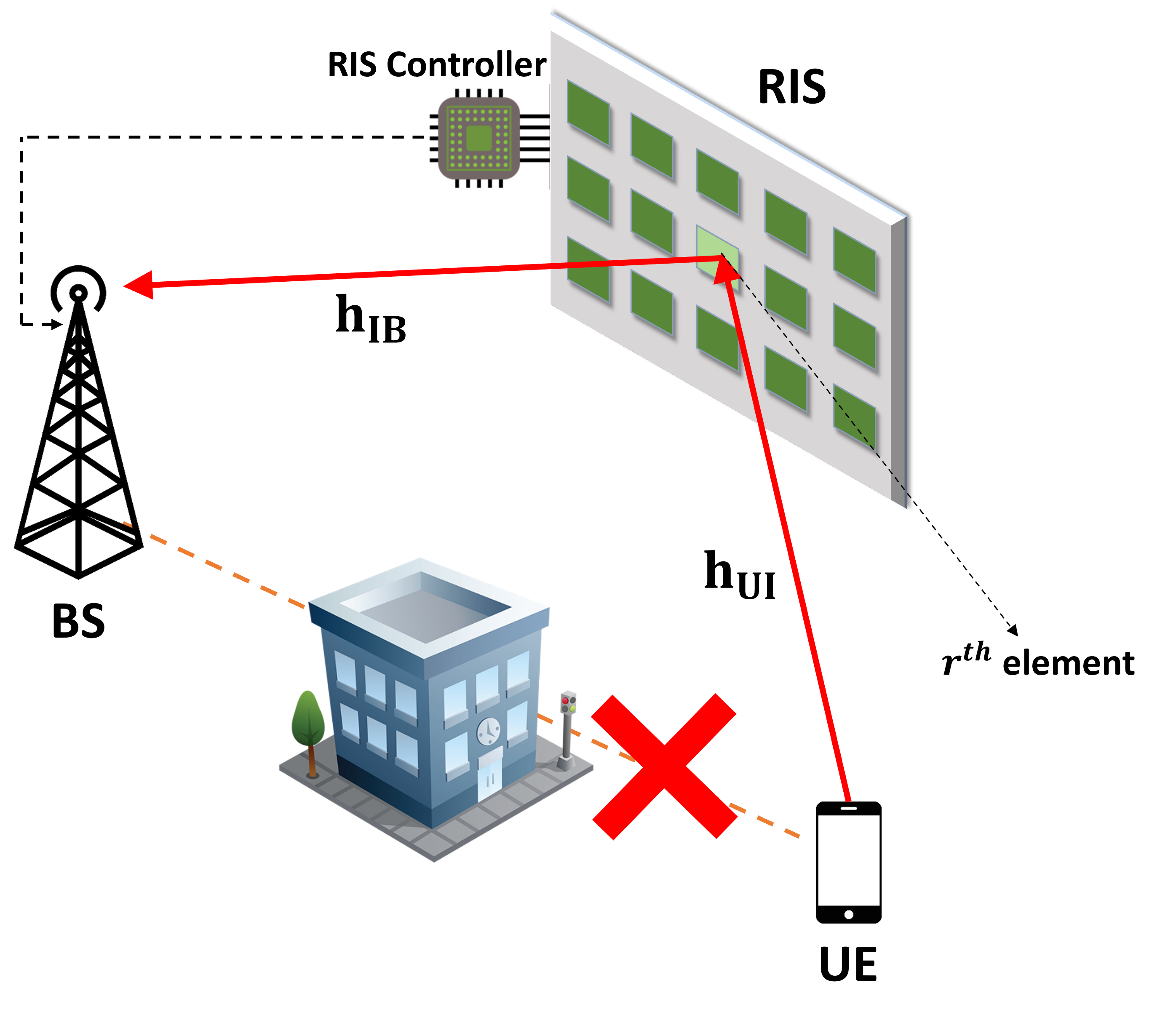}
     \centering
    \caption{RIS-assisted SC-FDE uplink SISO system with blocked direct UE–BS link. The BS controls the RIS via a dedicated controller.
    }
    \label{fig:system}
\end{figure}

We consider a single-input single-output (SISO) uplink communication system assisted by an RIS, as illustrated in Fig.~\ref{fig:system}. The RIS comprises $R$ passive reflecting elements, whose reflection coefficients are individually adjusted through a programmable controller at the base station (BS), which is assumed to have perfect channel state information (CSI). The reflection coefficient of the $r^{th}$ RIS element is defined as $\phi_r = a_r e^{j\theta_r}$, where $a_r$ represents the reflection amplitude and $\theta_r \in [0, 2\pi)$ is the adjustable phase chosen from a discrete set $\Theta$.  The wireless links between the user equipment (UE), the RIS, and the BS are modeled as frequency-selective multipath channels. Specifically, the channel impulse response (CIR) from the UE to the \(r^{\text{th}}\) RIS element is denoted by \(\mathbf{h}_{\mathrm{UI}}^{(r)} = [h_{\mathrm{UI}}^{(r)}(1),\, h_{\mathrm{UI}}^{(r)}(2),\, \ldots,\, h_{\mathrm{UI}}^{(r)}(L_{\mathrm{UI}})]^T\), and that from the \(r^{\text{th}}\) RIS element to the BS by \(\mathbf{h}_{\mathrm{IB}}^{(r)} = [h_{\mathrm{IB}}^{(r)}(1),\, h_{\mathrm{IB}}^{(r)}(2),\, \ldots,\, h_{\mathrm{IB}}^{(r)}(L_{\mathrm{IB}})]^T\), where \(L_{\mathrm{UI}}\) and \(L_{\mathrm{IB}}\) denote the number of taps for the UE-RIS and RIS-BS channels, respectively. Each component of \(\mathbf{h}_{\mathrm{UI}}^{(r)}\) and \(\mathbf{h}_{\mathrm{IB}}^{(r)}\) corresponds to the complex gain of the \(l^{\text{th}}\) multipath component. The end-to-end channel is constructed by convolving the UE–RIS and RIS–BS channels for each RIS element, weighted by its corresponding reflection coefficient, given by

\begin{equation}
    \mathbf{\tilde{h}} = \sum_{r=1}^{R} \left(\mathbf{h}_{\mathrm{UI}}^{(r)} * \mathbf{h}_{\mathrm{IB}}^{(r)}\right)\phi_r \in \mathbb{C}^{L \times 1},
\end{equation}
where \( * \) denotes discrete-time convolution, and \(L = L_{\mathrm{UI}} + L_{\mathrm{IB}} - 1\) is the length of the cascaded channel.

\subsection{Tap-Based RIS Phase Configuration}
RIS phase configuration can be designed to align the phases of individual multipath components in the end-to-end channel $\mathbf{\tilde{h}}$, enabling coherent combining at a desired delay tap. For each candidate tap index $\ell \in \{1, \dots, L\}$, the phase shift of the $r^{th}$ RIS element is configured to compensate for the phase of the $\ell^{th}$ tap as
\begin{equation}
   \theta_r^{(\ell)} = -\angle \left( [\mathbf{h}_{\mathrm{UI}}^{(r)} * \mathbf{h}_{\mathrm{IB}}^{(r)}](\ell) \right),
\end{equation}
where $[\cdot](\ell)$ denotes the $\ell^{th}$ tap of the cascaded CIR for the $r^{th}$ RIS path, and $\angle(\cdot)$ denotes the complex phase. The corresponding RIS reflection coefficient is then set as $\phi_r^{(\ell)} = \exp\left(j\theta_r^{(\ell)}\right)$.

Each value of $\ell$ corresponds to a distinct tap compensation method, such as first-tap alignment~\cite{over-the-air}, middle-tap alignment~\cite{tariq2025ris}, and maximum-gain tap selection~\cite{10843788}. In the latter case, phase compensation is computed for each candidate tap index $\ell$, and the tap that yields the maximum power is selected as
\begin{equation}
    \ell^* = \arg\max_{\ell} \left| \tilde{h}^{(\ell)}(\ell) \right|^2,
\end{equation}
where $\tilde{h}^{(\ell)}(\ell)$ denotes the $\ell^{th}$ tap of the cascaded channel obtained after compensating for tap $\ell$. This formulation enables a comparative analysis of different tap selections and their impact on the cascaded channel gain.

\subsection{Transmitter Design}
The UE maps \( b = N\log_2(M) \) information bits into an \( M \)-ary modulated vector \( \mathbf{x} = [x_1, \dots, x_N]^T \in \mathbb{C}^{N \times 1} \), where each symbol \( x_n \in \mathcal{X} \) is drawn from a complex \( M \)-ary constellation. To eliminate IBI, a CP of length \( L_{\mathrm{cp}} \geq L \) is appended to $\mathbf{x}$. Hence, the transmitted vector is expressed as
\begin{equation}
\mathbf{x}_{\mathrm{CP}} = [x(N - L_{\mathrm{cp}} + 1), \ldots, x(N), x(1), \ldots, x(N)]^T.
\end{equation}
At the receiver, the first \( L_{\mathrm{cp}} \) samples are discarded, and the remaining \( N \) samples are used for processing. Under the assumption \( N \geq L \), the linear convolution between the transmit signal and the channel can be accurately represented as a circular convolution. Accordingly, the input–output relation is given by
\begin{equation}
\label{eq:y}
    \mathbf{y} = \tilde{\mathbf{H}} \mathbf{x} + \mathbf{w},
\end{equation}
where \( \tilde{\mathbf{H}} \in \mathbb{C}^{N \times N} \) is a circulant matrix representing the end-to-end channel, and \( \mathbf{w} \sim \mathcal{CN}(0, \sigma_w^2 \mathbf{I}) \) denotes additive white Gaussian noise.
\subsection{Receiver Design}
In this work, detection performance is evaluated at the BS using two schemes: (i) frequency-domain MLD as a classical benchmarks, and (ii) the proposed hybrid quantum–classical detector applied to the considered system.

\textit{1) Frequency Domain MLD:} MLD provides optimal performance by minimizing the Euclidean distance between the received signal and all possible transmit sequences. Based on the signal model in~\eqref{eq:y}, the MLD estimate is obtained as
\begin{equation}
    \hat{\mathbf{x}} = \arg\min_{\mathbf{x} \in \mathcal{X}^N} \left\| \mathbf{y} - \tilde{\mathbf{H}} \mathbf{x} \right\|^2.
\end{equation}
Given that \( \tilde{\mathbf{H}} \) is circulant, it can be diagonalized as \( \tilde{\mathbf{H}} = \mathbf{Q}^{\dagger} \boldsymbol{\Lambda} \mathbf{Q} \), where \( \mathbf{Q} \) is the unitary DFT matrix, \( (\cdot)^{\dagger} \) denotes the Hermitian operator, and  \( \boldsymbol{\Lambda} \) is a diagonal matrix with entries corresponding to the frequency-domain coefficients of \( \tilde{\mathbf{h}} \). Applying this transformation to both sides of the signal model yields the frequency-domain formulation of the MLD problem, which is given by
\begin{equation}
\label{eq:MLD}
    \hat{\mathbf{x}} = \arg\min_{\mathbf{x}_f \in \mathcal{X}_f^N} \left\| \mathbf{y}_f - \boldsymbol{\Lambda} \mathbf{x}_f \right\|^2,
\end{equation}
where \( \mathbf{x}_f = \mathbf{Q} \mathbf{x} \) and \( \mathbf{y}_f = \mathbf{Q} \mathbf{y} \) are the frequency-domain transmit and received vectors, respectively.

While MLD provides the minimum achievable error probability, its computational complexity scales exponentially with both \( M \) and \( N \), making it impractical for wideband systems or large constellations.

\textit{2) Hybrid quantum–classical Detection:} This scheme employs a quantum-assisted detector based on the GAS algorithm, which uses amplitude amplification to identify the most likely transmitted symbol vector efficiently. The detection problem is reformulated as a QUBO task, enabling the exploitation of quantum parallelism for efficient search. The GAS-based approach achieves near-optimal performance, closely approaching that of classical maximum likelihood detection, while offering a quadratic reduction in query complexity. Detailed implementation aspects are discussed in the following sections.
\section{QUBO Formulation}
\label{Implementation}
The frequency-domain MLD problem in~\eqref{eq:MLD} is reformulated as a QUBO formulation, which is compatible with quantum search and optimization algorithms. This transformation enables efficient implementation on quantum hardware, including annealers and gate-based solvers. Leveraging the diagonal structure of $\boldsymbol{\Lambda}$ eliminates the need for full matrix multiplications in the QUBO construction and hence reduces the classical preprocessing complexity.
 This simplification is particularly beneficial for large block lengths, enabling scalable QUBO generation. To enable QUBO-based detection, the MLD cost function is reformulated in terms of the time-domain variable \( \mathbf{x} \) by substituting \( \mathbf{x}_f = \mathbf{Q} \mathbf{x} \) into the frequency-domain formulation in~\eqref{eq:MLD}, yielding
\begin{align}
E(\mathbf{x}) 
&= \left\| \mathbf{y}_f - \boldsymbol{\Lambda} \mathbf{Q} \mathbf{x} \right\|^2 \nonumber \\
&= \mathbf{x}^{\dagger} \mathbf{Q}^{\dagger} \boldsymbol{\Lambda}^{\dagger} \boldsymbol{\Lambda} \mathbf{Q} \mathbf{x} 
- 2 \Re \left\{ \mathbf{x}^{\dagger} \mathbf{Q}^{\dagger} \boldsymbol{\Lambda}^{\dagger} \mathbf{y}_f \right\} 
+ \mathbf{y}_f^{\dagger} \mathbf{y}_f.
\label{eq:cost_dft}
\end{align}
The constant term \( \mathbf{y}_f^{\dagger} \mathbf{y}_f \), which is independent of \( \mathbf{x} \), can be omitted from the optimization, hence, the cost function reduces to
\begin{equation}
\label{eq:1QUBO}
E(\mathbf{x}) = 
\mathbf{x}^{\dagger} \mathbf{Q}^{\dagger} \boldsymbol{\Lambda}^{\dagger} \boldsymbol{\Lambda} \mathbf{Q} \mathbf{x} 
- 2 \Re \left\{ \mathbf{x}^{\dagger} \mathbf{Q}^{\dagger} \boldsymbol{\Lambda}^{\dagger} \mathbf{y}_f \right\} + \text{const}.
\end{equation}
Defining
\begin{equation}
\mathbf{Q}_{\mathrm{BPSK}} = \mathbf{Q}^{\dagger} \boldsymbol{\Lambda}^{\dagger} \boldsymbol{\Lambda} \mathbf{Q}, \quad 
\mathbf{c}_{\mathrm{BPSK}} = -2\Re\left\{ \mathbf{Q}^{\dagger} \boldsymbol{\Lambda}^{\dagger} \mathbf{y}_f \right\},
\end{equation}
the cost function in \eqref{eq:cost_dft} can be expressed as
\begin{equation}
\label{eq:cost}
E(\mathbf{x}) = \mathbf{x}^{\dagger} \mathbf{Q}_{\mathrm{BPSK}} \mathbf{x} + \mathbf{c}_{\mathrm{BPSK}}^\top \mathbf{x} + \text{const}.
\end{equation}
For BPSK modulation, where each transmitted symbol \( x_i \in \{-1, +1\} \), the symbols are equivalently expressed in terms of binary variables \( b_i \in \{0, 1\} \) using the mapping \( x_i = 2b_i - 1 \). Substituting this relation into (\ref{eq:cost}) yields
\begin{align}
\label{eq:qubo_expand}
\small
E(\mathbf{b}) &= (2\mathbf{b} - \mathbf{1})^{\dagger} \mathbf{Q}_{\mathrm{BPSK}} (2\mathbf{b} - \mathbf{1}) 
+ (2\mathbf{b} - \mathbf{1})^\top \mathbf{c}_{\mathrm{BPSK}} \nonumber \\
&= 4\mathbf{b}^\top \mathbf{Q}_{\mathrm{BPSK}} \mathbf{b} 
+ (-4\mathbf{Q}_{\mathrm{BPSK}}\mathbf{1} + 2\mathbf{c}_{\mathrm{BPSK}})^\top \mathbf{b} 
+ \text{const}.
\end{align}
Accordingly, the cost function takes the standard QUBO form given by
\begin{equation}
\label{eq:final_QUBO}
E(\mathbf{b}) = \mathbf{b}^\top \mathbf{Q}' \mathbf{b} + \mathbf{c}'^\top \mathbf{b} + \text{const},
\end{equation}
where \( \mathbf{Q}' = 4\mathbf{Q}_{\mathrm{BPSK}} \) and \( \mathbf{c}' = -4\mathbf{Q}_{\mathrm{BPSK}}\mathbf{1} + 2\mathbf{c}_{\mathrm{BPSK}} \). The resulting binary quadratic form is directly compatible with quantum optimization algorithms such as GAS. 
\section{Grover Adaptive Search}
\label{sec:GAS}
\begin{algorithm}[t]
\caption{Grover Adaptive Search~\cite{Gilliam2021groveradaptive}}
\label{alg:real_valued_gas}
\begin{algorithmic}[1]
\REQUIRE $E : \mathbb{B}^n \rightarrow \mathbb{R}$, $\lambda = 8/7$
\ENSURE $\mathbf{b}$
\STATE Sample $\mathbf{b}_0 \in \mathbb{B}^n$ and set $y_0 = E(\mathbf{b}_0)$
\STATE Initialize $k = 1$ and $i = 0$
\REPEAT
    \STATE Sample $L_i \in \{0, 1, \ldots, \lceil k - 1 \rceil\}$
    \STATE Execute $G^{L_i} A_{y_i} \ket{0}^{\otimes(n+m)}$ and measure $\mathbf{b}$
    \STATE Compute $E(\mathbf{b})$ classically
    \IF{$E(\mathbf{b}) < y_i$}
        \STATE $\mathbf{b}_{i+1} = \mathbf{b}$, $y_{i+1} = E(\mathbf{b})$, $k = 1$
    \ELSE
        \STATE $\mathbf{b}_{i+1} = \mathbf{b}_i$, $y_{i+1} = y_i$, $k = \min(\lambda k, \sqrt{2^n})$
    \ENDIF
    \STATE $i \leftarrow i + 1$
\UNTIL termination condition met
\end{algorithmic}
\end{algorithm}
GAS is an iterative quantum algorithm that adaptively minimizes objective functions by dynamically updating a cost threshold. Unlike Grover variants that operate with a fixed threshold or rely on pre-defined bounds, GAS adjusts its threshold based on observed outcomes, enabling efficient exploration of unknown or real-valued cost landscapes. At each step, the current threshold $\gamma_i$ acts as a benchmark, and candidate states with cost $E(\mathbf b)$ below $\gamma_i$ are considered improvements. This motivates encoding the shifted difference $E(\mathbf b)-\gamma_i$ into the quantum state. Each iteration applies a Grover operator given by
\begin{equation}
\label{eq:qrover}
    G = A_{\gamma_i} D A_{\gamma_i}^{\dagger} O_{\gamma_i},
\end{equation}
where $A_{\gamma_i}$ denotes the state-preparation circuit that embeds the cost function relative to the threshold $\gamma_i$, 
$O_{\gamma_i}$ is the oracle that flips the phase of states with costs below $\gamma_i$, and $D$ is the diffusion operator that performs inversion about the mean. The number of Grover calls $L_i$ is sampled uniformly from $\{0,1,\ldots,\lceil k_i-1\rceil\}$, with $k_i$ serving as an adaptive control parameter. After measurement, if $E(\mathbf b) < \gamma_i$, the threshold is updated to $\gamma_{i+1} = E(\mathbf b)$ and the control parameter is reset to $k_{i+1} = 1$; otherwise, $k_{i+1}$ is scaled by a factor $\lambda > 1$ (typically $\lambda = 8/7$) until it reaches $O(\sqrt{2^n})$, encouraging broader exploration in subsequent iterations.

\section{Adaptive Threshold Initialization}
\label{sec:threshold}

The performance of GAS depends critically on the choice of the initial threshold, which determines the set of candidate solutions amplified during the search. In its original formulation, the threshold is derived from a randomly sampled candidate, often resulting in slow convergence due to poor initial guidance. In this work, a frequency-domain MMSE equalizer is employed to initialize the GAS threshold. This choice accelerates convergence by reducing the number of required Grover iterations, thereby yielding shallower circuits that are more robust to noise. Unlike prior GAS studies that relied on spatial-domain MMSE detectors~\cite{6730885,10044091}, the proposed approach exploits the fact that the cyclic prefix renders $\tilde{\mathbf{H}}$ circulant, which diagonalizes under the DFT. Consequently, MMSE equalization reduces to simple element-wise multiplications in the frequency domain, avoiding the large-scale matrix inversions required in the time domain. Frequency equalization is  performed with the diagonal MMSE filter matrix $\mathbf{\Phi} \in \mathbb{C}^{N \times N}$, whose entries are
\begin{equation}
    \Phi(i,i) = \frac{\Lambda^*(i,i)}{|\Lambda(i,i)|^2 + \sigma_w^2 / \sigma_x^2},
\end{equation}
where $\sigma_x^2$ and $\sigma_w^2$ denote the symbol and noise variances, respectively, and $\Lambda(i,i)$ is the $i^{th}$ diagonal element of the diagonalized channel matrix $\boldsymbol{\Lambda}$. The equalized signal is then transformed back to the time domain as
\begin{equation}
    \mathbf{d} = \mathbf{Q}^{\dagger} \mathbf{\Phi} \mathbf{y}_f.
\end{equation}
A hard-decision estimate $\bar{\mathbf{x}}_0 \in \mathcal{X}^N$ is obtained by mapping $\mathbf{d}$ to the nearest constellation points, and the corresponding QUBO objective value is
\begin{equation}
    \bar{y}_0 = \left\| \mathbf{y} - \tilde{\mathbf{H}} \bar{\mathbf{x}}_0 \right\|^2.
\end{equation}
Since $\bar{\mathbf{x}}_0$ is feasible but generally suboptimal, $\bar{y}_0$ provides a conservative upper bound on the unknown ML cost. Initializing GAS with this threshold keeps the ML solution within the search space while reducing Grover iterations and circuit depth, improving robustness to noise on NISQ hardware.

\section{Grover-based Quantum Circuit}
\label{sec:Circuit}
This section details the construction of the quantum circuit implementing GAS, including the quantum registers, the state-preparation circuit, the oracle, and the diffusion operator.

\subsection{Quantum Register Structure}
GAS circuit comprises two main registers,  \emph{key register} and \emph{value register}.
\begin{itemize}
    \item \emph{Key register:} An $n$-qubit register encodes the set of candidate transmitted vectors, where $n = N \log_2 M$. Each basis state $\ket{\mathbf{b}} \in \{0,1\}^n$ represents a distinct hypothesis over the $M^N$ possible symbol vectors. For binary modulation (e.g., BPSK), $n = N$ and each bit $b_i$ maps to a symbol via $b_i \mapsto (-1)^{1 - b_i}$.

    \item \emph{Value register:} An $m$-qubit register, denoted by $\{z_j\}_{j=0}^{m-1}$, is used to represent the cost function value $E(\mathbf{b})$, derived from the QUBO formulation in (\ref{eq:final_QUBO}). This register is encoded in two's complement format to accommodate both positive and negative values. To enable comparisons between $E(\mathbf{b})$ and the dynamic threshold $\gamma_i$ during GAS, the register must support all possible difference values $\xi = E(\mathbf{b}) - \gamma_i$. Defining $\mathcal{E}_{\max} = \max_{\mathbf{b}} E(\mathbf{b})$ and $\mathcal{E}_{\min} = \min_{\mathbf{b}} E(\mathbf{b})$, the admissible range of differences is
    \begin{equation}
        \xi \in [\mathcal{E}_{\min} - \mathcal{E}_{\max},\; \mathcal{E}_{\max} - \mathcal{E}_{\min}].
    \end{equation}
    Since an $m$-qubit two’s complement register encodes integers in the range $[-2^{m-1},\,2^{m-1}-1]$, the bitwidth $m$ must satisfy
    \begin{equation}
        \label{m_selection}
        m > \left\lceil \log_2 \left( \mathcal{E}_{\max} - \mathcal{E}_{\min} + 1 \right) \right\rceil .
    \end{equation}
    For example, if $\mathcal{E}_{\max} = 11$ and $\mathcal{E}_{\min} = -5$, then the maximum magnitude of any difference is $16$, which requires $m = 5$ qubits since a 5-bit two’s complement register spans $[-16, 15]$. In practice, $m$ is chosen conservatively to ensure robustness across channel realizations and system configurations.
\end{itemize}

This register structure enables parallel representation of all candidate symbol vectors in the \emph{key register}, while reserving the \emph{value register} for coherent encoding of the cost function. Together, they provide the foundation for the state-preparation circuit $A_{\gamma_i}$ described next.

\subsection{State Preparation Circuit}

The state-preparation circuit $A_{\gamma_i}$ operates on the defined quantum registers to coherently embed the MLD cost landscape into the quantum state. Its operation comprises three stages: (i) creation of an equal superposition via Hadamard gates, (ii) coherent phase encoding of the cost function shifted by the threshold, and (iii) application of the inverse quantum Fourier transform (IQFT).

\subsubsection{Hadamard Layer}
The state preparation begins by creating a uniform superposition over all candidate symbol vectors. Starting from the ground state $\ket{0}^{\otimes n}\ket{0}^{\otimes m}$, Hadamard gates are applied to the qubits encoding the candidate bitstrings, while the value register remains initialized at $\ket{0}^{\otimes m}$. The resulting state is given by
\begin{equation}
\small
|\psi_0\rangle
= (H^{\otimes n}\!\otimes I^{\otimes m})\,|0\rangle^{\otimes n}|0\rangle^{\otimes m}
= \frac{1}{\sqrt{2^n}}\sum_{\mathbf{b}\in\{0,1\}^n} 
   \ket{\mathbf{b}}\,\ket{0}^{\otimes m},
   \label{eq:first}
\end{equation}
where $I^{\otimes m}$ denotes the $m$-qubit identity operator acting on the \emph{value register}, and $\ket{\mathbf{b}}$ represents a candidate bitstring in the \emph{key register}. At this stage, the quantum system coherently represents all candidate hypotheses in parallel.

\subsubsection{Phase Encoding}
\label{Phase_enc}
GAS conventional algorithm is designed for unstructured search tasks, where an
oracle marks solution states by flipping their phase~\cite{10044091}. In contrast, the MLD problem reformulated as a QUBO is expressed through a real-valued cost function, which such oracles cannot directly process. To enable cost-aware quantum search, the cost function is encoded into the
quantum state via controlled phase rotations, implemented by a unitary operator
$U(\theta)$. Rather than
storing cost values explicitly in a register, the cost $E(\mathbf{b})$ is
encoded into the relative phase of each basis state $\ket{\mathbf{b}}$,
enabling threshold-based comparisons through quantum interference. Gilliam
\textit{et al.}~\cite{Gilliam2021groveradaptive} proposed two strategies for
phase-based cost encoding: (i) integer approximation and (ii) direct real-valued
encoding.

\paragraph*{Integer Approximation}
In this approach, the objective function is uniformly rescaled so that all
real-valued coefficients become integers.
The overall optimization landscape is preserved up to
this scaling, but the dynamic range of values grows with the scaling factor.
Consequently, the number of qubits in the \emph{value register} must also grow to prevent overflow, otherwise the representation becomes inaccurate. A larger $m$ increases circuit depth and the number of multi-controlled $R_z$ gates, which reduces scalability on noisy hardware.

\paragraph*{Direct Real-Valued Encoding}
This approach avoids the limitations of integer rescaling by directly mapping real-valued costs into phase rotations, thereby eliminating quantization. Each basis state $\ket{\mathbf{b}}$ acquires a phase proportional to its shifted cost $E(\mathbf{b}) - \gamma_i$, with rotation angle
\begin{equation}
    \theta(\mathbf{b}) = \frac{2\pi}{2^m}\,\big(E(\mathbf{b}) - \gamma_i\big),
\end{equation}
where the factor $2\pi/2^m$ ensures resolvable phase differences across the \emph{value register} when processed by the IQFT. The encoding is implemented by a sequence of controlled and multi-controlled $R_z$ gates, where qubits from the \emph{key register} act as controls, and qubits from the \emph{value register} receive binary-weighted phase shifts.

Each coefficient in (\ref{eq:1QUBO}) is mapped onto the
\emph{value register} by decomposing it into binary-weighted phase rotations.
In the binary expansion of the \emph{value register}, the $j^{th}$ qubit corresponds to
weight $2^j$, with elementary rotation defined as
\begin{equation}
    \theta_j = 2^j \cdot \tfrac{2\pi}{2^m}.
\end{equation}
A coefficient $c$ is therefore mapped to a total phase
$\tfrac{2\pi}{2^m}c$, implemented by distributing its binary expansion across
the appropriate $z_j$ qubits. Linear terms use single-controlled rotations,
quadratic terms use double-controlled rotations, and the constant threshold
$-\gamma_i$ is applied via unconditional rotations.

As an illustration, consider $m=3$, which gives
$\theta_0 = \pi/4$, $\theta_1 = \pi/2$, and $\theta_2 = \pi$. For a linear term
$\alpha b_0$ with $\alpha=3$, the total phase contribution is
$\tfrac{2\pi}{8}\cdot 3 = 3\pi/4$, which decomposes into $\pi/2 + \pi/4$.
Accordingly, $b_0$ controls an $R_z(\pi/4)$ on $z_0$ and an $R_z(\pi/2)$ on
$z_1$, while $z_2$ remains unaffected in this case.
\begin{figure*}[t] 
    \centering
    \includegraphics[width=1\linewidth]{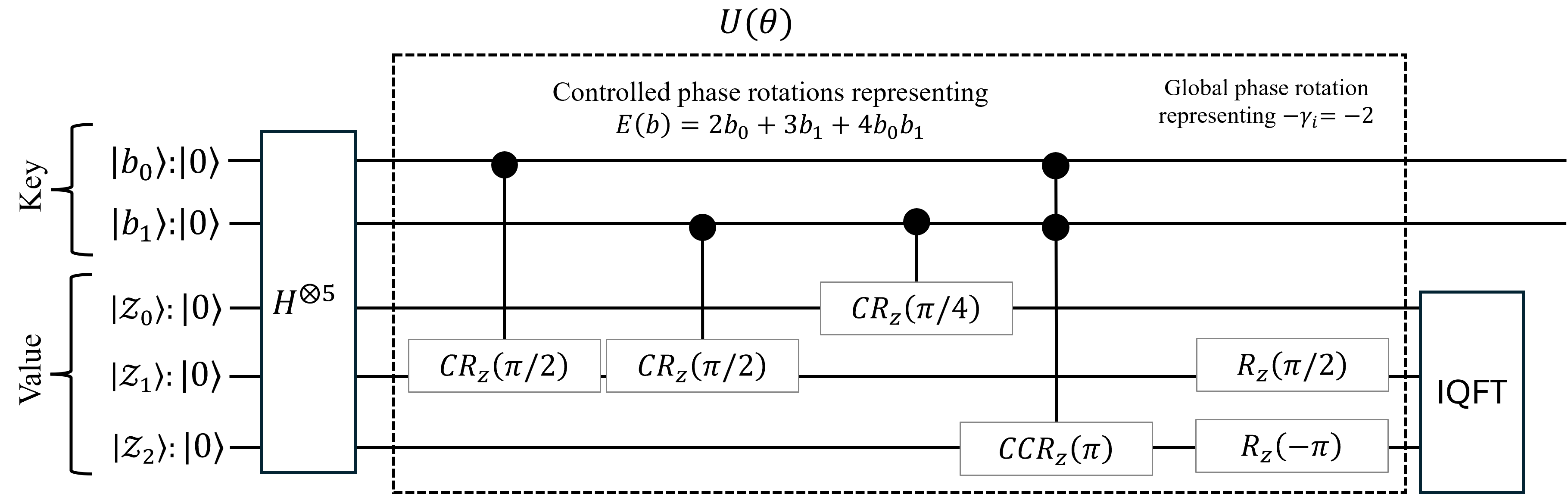}
   \caption{Quantum circuit of the GAS state-preparation stage. 
The example cost is $E(b) = 2b_0 + 3b_1 + 4b_0 b_1$ with threshold $\gamma_i = 2$. 
A value register of $m=3$ qubits encodes the cost in binary (weights $1,2,4$), while a key register of $n=2$ qubits holds candidate bitstrings $\mathbf b=(b_0,b_1)$. 
Controlled rotations implement the cost terms, unconditional rotations encode subtraction of $\gamma_i$ in two’s-complement form, and the final IQFT maps accumulated phases into measurable amplitudes.}

    \label{fig:circuit}
\end{figure*}

\subsubsection{Inverse Quantum Fourier Transform (IQFT)}
\label{IQFT}
The final step of state preparation applies the IQFT to the \emph{value register}, transforming the accumulated phase rotations into measurable amplitudes. Conceptually, the overall state-preparation operator is represented as
\begin{equation}
\label{eq:A_def}
    A_{\gamma_i} \ket{0}^{\otimes n} \ket{0}^{\otimes m} 
    = \frac{1}{\sqrt{2^n}} \sum_{\mathbf{b} \in \{0,1\}^n} 
      \ket{\mathbf{b}} \ket{E(\mathbf{b}) - \gamma_i},
\end{equation}
which illustrates the intended entanglement between candidate bitstrings and their shifted costs. In practice, for direct real-valued encoding the post-IQFT measurement statistics follow the Fej\'er distribution~\cite{10044091}, with probability mass concentrated near the integers closest to the true cost. The corresponding quantum state is given by
\begin{equation}
U_{\text{Fej\'er}}(\theta)\ket{0}^{\otimes m}
    = \sum_{l=0}^{2^m - 1} \big\langle g(\theta),\, g(2\pi l / 2^m) \big\rangle \ket{l},
\end{equation}
where $g(\theta) = [1, e^{i\theta}, \dots, e^{i(2^m - 1)\theta}]/\sqrt{2^m}$.
Because of this spread, direct measurement of the \emph{value register} can 
misestimate $E(\mathbf b)$, leading to premature threshold updates that may 
discard valid candidates. To avoid this, after diffusion and measurement the 
cost is recomputed classically from the observed $\mathbf b$, ensuring exact 
objective evaluation while still leveraging the quantum search. This hybrid 
step preserves both robustness and convergence reliability.

To illustrate the state-preparation process, Fig.~\ref{fig:circuit} shows an example for a two-variable QUBO. The terms $2b_0$, $3b_1$, and 
$4b_0b_1$ are encoded by binary-weighted controlled rotations on the 
\emph{value register}, while the global rotation applies the threshold shift 
$-\gamma_i$. The resulting unitary $U(\theta)$ implements the state-preparation 
block whose output is subsequently processed by the IQFT for cost evaluation.

\begin{figure}
\includegraphics[width=1\linewidth]{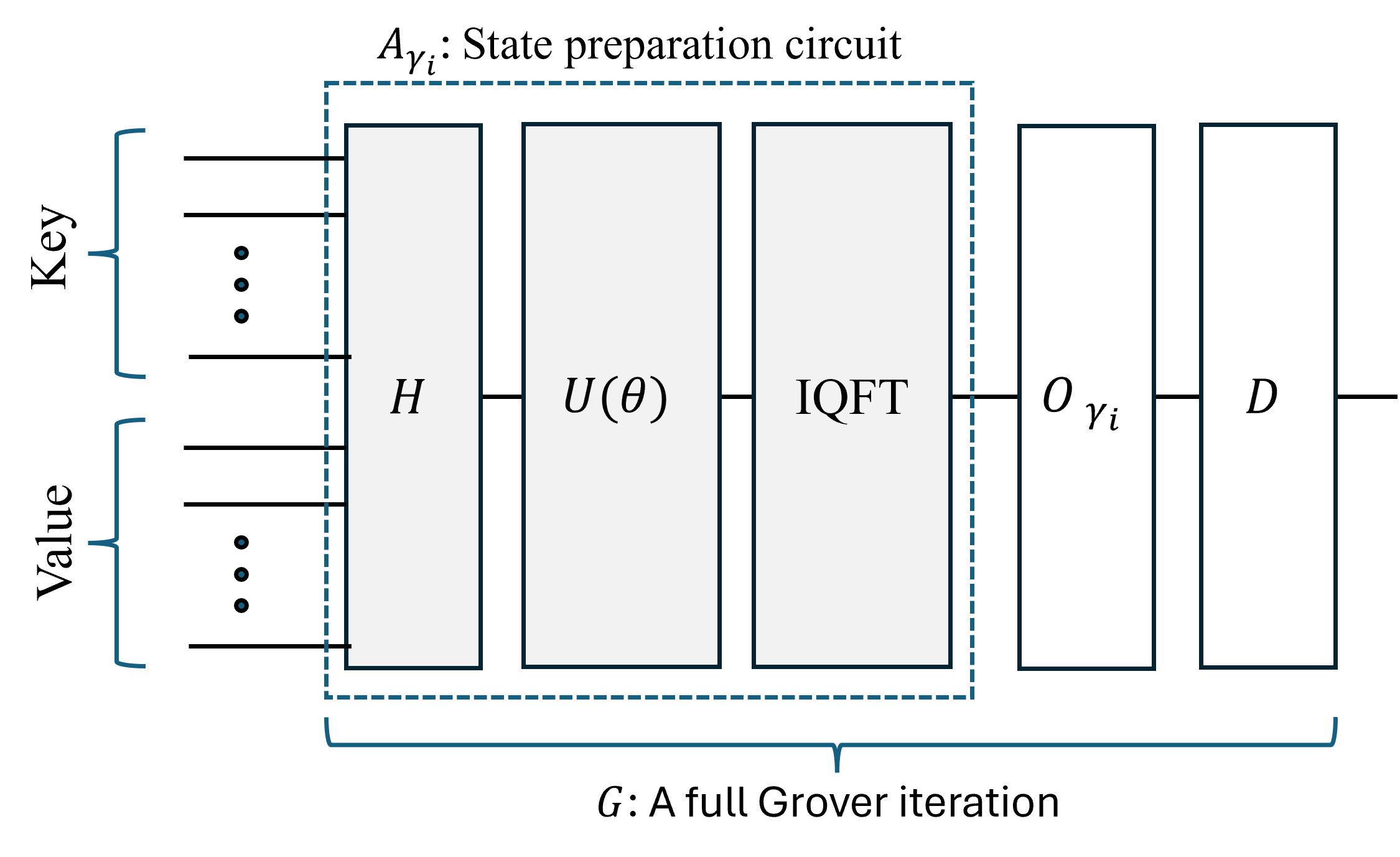}
    \caption{GAS circuit showing state preparation $A_{\gamma_{i}}$, oracle $O$, and diffusion operator $D$, iterated to amplify low-cost solutions based on threshold $\gamma_i$.}
    
    \label{fig:grover}
\end{figure}

\subsection{Oracle and Diffusion Operators}

After state preparation, the Grover-based search phase is executed through the oracle and diffusion operators. These components identify low-cost candidate solutions and iteratively amplify their probability amplitudes, increasing the likelihood of measuring optimal symbol vectors.

\subsubsection{Oracle Operator}

The oracle marks candidate bit strings $\mathbf{b}$ whose shifted cost is negative, i.e., $E(\mathbf b) - \gamma_i < 0$.  
Let $F_{\gamma_i}$ denote the \emph{compute-only} encoder, which evaluates the shifted cost on the \emph{value register} while leaving the \emph{key register} unchanged. Hence, $F_{\gamma_i}$ can be defined as
\begin{equation}
F_{\gamma_i}\,|\mathbf b\rangle|0\rangle^{\otimes m} 
= |\mathbf b\rangle\,|E(\mathbf b)-\gamma_i\rangle.
\end{equation}
This block relates to the state-preparation operator $A_{\gamma_i}$ through
\begin{equation}
A_{\gamma_i} = (H^{\otimes n} \otimes I^{\otimes m})\,F_{\gamma_i}, \quad
F_{\gamma_i} = (H^{\otimes n} \otimes I^{\otimes m})^{\dagger} A_{\gamma_i}.
\end{equation}  
A single application of $H^{\otimes n}$ prepares the uniform superposition, given by
\begin{equation}
|s\rangle = 2^{-n/2} \sum_{\mathbf b \in \{0,1\}^n} |\mathbf b\rangle,
\end{equation}
after which $F_{\gamma_i}$ is invoked in the compute–mark–uncompute sequence of each Grover iteration. Following the IQFT, the \emph{value register} holds $E(\mathbf b) - \gamma_i$ in two’s complement representation. Negativity is indicated by the most significant bit (MSB), with MSB = 1 marking a phase flip, which can be expressed as
\begin{equation}
|\mathbf b\rangle|E(\mathbf b)-\gamma_i\rangle \xrightarrow{\text{Oracle}}
\begin{cases}
-\,|\mathbf b\rangle|E(\mathbf b)-\gamma_i\rangle, & E(\mathbf b) < \gamma_i,\\[2pt]
\phantom{-}\,|\mathbf b\rangle|E(\mathbf b)-\gamma_i\rangle, & \text{otherwise}.
\end{cases}
\end{equation}
The marking operation is realized by applying a single $Z$ gate on the MSB, given by
\begin{equation}
O_{\gamma_i} = F_{\gamma_i}^{\dagger}\, Z\, F_{\gamma_i}.
\end{equation}
This design avoids intermediate measurements, reduces circuit overhead, and is adequate for the numerical experiments in this work.

\subsubsection{Diffusion Operator}

The diffusion operator performs an inversion about the mean amplitude of the \emph{key register} states, thereby amplifying those flagged by the oracle. For an \(n\)-qubit \emph{key register}, it is expressed as
\begin{equation}
\label{eq:diff}
    D = 2 \ket{\psi_0}\bra{\psi_0} - I,
\end{equation}
where $\ket{\psi_0}$ is the uniform superposition state given in (\ref{eq:first}). This transformation increases the amplitude of low-cost candidates while suppressing others, progressively steering the system toward optimal solutions with each iteration. The complete circuit for a single GAS iteration is illustrated in Fig.~\ref{fig:grover}.

\section{Quantum Resources}
\label{sec:resource}

This section quantifies the quantum resources required by the proposed GAS-based detector, beginning with its query complexity and followed by register widths and gate counts. Unless stated otherwise, gate counts are reported at the \emph{block} level, where controlled rotations ($CR_z$ and $CCR_z$) are treated as native primitives and the $m$-qubit IQFT is counted as a single block. This convention isolates the intrinsic algorithmic cost from hardware-specific decompositions into one- and two-qubit gates, which can be derived as needed without affecting the asymptotic conclusions.

\subsection{Query Complexity}
\label{sec:query_complexity}

The query complexity of GAS quantifies the number of times the cost function is accessed during the algorithm and consists of two components, a \emph{classical} part from explicit evaluations of $E(\mathbf b)$ and a \emph{quantum} part from repeated applications of the Grover operator.

On the classical side, each iteration ends with a measurement of a candidate $\mathbf b$, whose cost $E(\mathbf b)$ is recomputed to update the adaptive threshold $\gamma_i$. Including the initial evaluation used to set $\gamma_0$, the total number of classical queries after $T$ iterations is $T+1$.

On the quantum side, complexity is determined by the number of applications of the Grover operator in~(\ref{eq:qrover}), which comprises state preparation, the oracle, and diffusion. In the $i^{th}$ iteration, the randomized schedule applies $G^{L_i}$, meaning $L_i$ sequential calls of $G$. After $T$ iterations, the total number of Grover calls is $\sum_{i=1}^T L_i$. When $L_i$ is sampled uniformly from $\{0,1,\ldots,\lceil k_i - 1\rceil\}$, the expected number of calls per iteration is $\mathbb{E}[L_i] = \lceil k_i - 1 \rceil/2$, giving an expected quantum query complexity of $\sum_{i=1}^T \lceil k_i - 1 \rceil / 2$.

Since $k_i$ resets to $1$ when an improved solution is found and scales by $\lambda > 1$ otherwise, it grows adaptively and can reach as large as the Grover-optimal bound $O(\sqrt{M^N})$. This ensures that GAS explores the amplitude space without overshooting while preserving Grover’s quadratic speedup. Consequently, the expected quantum query complexity is $O(\sqrt{M^N})$, compared to the $O(M^N)$ evaluations required for classical exhaustive search.

\subsection{Register Widths}
For BPSK modulation, the \emph{key register}, which encodes candidate symbol sequences, requires one qubit per transmitted symbol, i.e., $n = N$.  
The \emph{value register} consists of $m$ qubits that represent the shifted cost $E(\mathbf{b}) - \gamma_i$ in two’s-complement form, with $m$ chosen according to the dynamic range of QUBO energies as specified in~\eqref{m_selection}.  
The total qubit requirement is thus $n+m$.

\subsection{Quantum Gate Counts}
\subsubsection{Gate Counts for $A_{\gamma_i}$}
The gate complexity of the state-preparation block $A_{\gamma_i}$ follows directly from the expanded QUBO cost function in~(\ref{eq:final_QUBO}). Linear monomials arise from the diagonal entries of $\mathbf Q'$ and the vector $\mathbf c'$. Since $b_i^2 = b_i$ for binary variables, each key qubit contributes one linear term, giving $N_{\mathrm{lin}} = n$. Quadratic monomials correspond to off-diagonal entries of $\mathbf Q'$, which capture pairwise interference between transmitted bits. In an ISI channel with memory length $L-1$, each bit interacts with at most $L-1$ neighbors. Symmetry ensures that only unique pairs are retained, resulting in $N_{\mathrm{quad}} = n(L-1)$ quadratic terms.

In the quantum circuit, each linear monomial is mapped to $m$ single-controlled rotations ($CR_z$), for a total of $nm$ gates. Each quadratic monomial maps to $m$ double-controlled rotations ($CCR_z$), giving $n(L-1)m$ gates overall. In addition, $A_{\gamma_i}$ applies $n+m$ Hadamard gates, $m$ unconditional $R_z$ gates to encode subtraction of $\gamma_i$ for MSB-based oracle detection, and one $m$-qubit IQFT. The gate counts for $A_{\gamma_i}$ with BPSK are summarized in Table~\ref{tab:gate_counts}.

\begin{table}[h]
\centering
\centering
\normalsize
\renewcommand{\arraystretch}{1.1}
\caption{Quantum gate counts per $A_{\gamma_i}$ for BPSK.}
\begin{tabular}{|l|c|}
\hline
Gate type & Count in $A_{\gamma_i}$ \\
\hline
$H$ & $n+m$ \\
\hline
$R_z$ & $m$ \\
\hline
$CR_z$ & $nm$ \\
\hline
$CCR_z$ & $n(L-1)m$ \\
\hline
$\beta$-controlled $R_z$ ($\beta \geq 3$) & $0$ \\
\hline
IQFT & $1$ \\
\hline
\end{tabular}
\label{tab:gate_counts}
\end{table}
\subsubsection{Overall Gate Count per Iteration}
In iteration $i$, the GAS circuit executes $G^{L_i} A_{\gamma_i}$.  
The initial call to $A_{\gamma_i}$ prepares the uniform superposition and encodes the cost from the all-zero state before Grover iterations begin. Each Grover operator $G$ then applies both $A_{\gamma_i}$ and its inverse $A_{\gamma_i}^\dagger$, so the state-preparation block is invoked twice per Grover step.  
Including the initial preparation, $A_{\gamma_i}$ appears a total of $2L_i+1$ times in iteration $i$. The oracle $O_{\gamma_i}$ contributes one $Z$ gate on the most-significant value qubit per Grover step, while the diffusion operator $D$ adds $L_i$ applications of the $n$-qubit inversion-about-the-mean.  
Hence, the overall per-iteration gate cost scales linearly with $L_i$, dominated by repeated state-preparation and its inverse, with lighter contributions from the oracle and diffusion.

\section{Quantum Noise Modeling}
\label{sec:quantum_noise}

To assess the robustness of the proposed quantum-assisted detection scheme under realistic hardware constraints, we simulate depolarizing noise \cite{berthusen2025toward} and readout noise \cite{maciejewski2020mitigation}, which are widely used abstractions for modeling stochastic gate errors and measurement inaccuracies in NISQ devices. These two noise processes capture the dominant effects of gate and measurement imperfections in current superconducting architectures and are particularly relevant for Grover-style algorithms, where shallow-depth circuits are limited mainly by such errors rather than long-time decoherence.

Although quantum hardware exhibits a wider range of physical noise processes, such as amplitude damping, phase decoherence, and thermal relaxation, these more detailed models require device-specific calibration parameters. In contrast, depolarizing and readout noise provide platform-agnostic abstractions that are widely adopted for algorithm-level benchmarking in the NISQ regime~\cite{pokharel2024better}. They balance fidelity and generality, capturing the essential impact of gate and measurement errors without relying on hardware-specific details.

\subsection{Depolarizing Noise}
Depolarizing noise is a standard abstraction for modeling stochastic gate-level imperfections in quantum circuits. It approximates the cumulative effect of various physical noise processes by replacing the qubit state with the maximally mixed state with probability \(p\). This model is widely used for algorithm-level benchmarking owing to its platform-independent generality and analytic tractability. For single-qubit gates, the depolarizing noise model acts on a density matrix \(\rho\) as \cite{alma991000408589707456}
\begin{equation}
\mathcal{E}_{\mathrm{dep}}(\rho) = (1-p)\rho + \frac{p}{3}\big(X\rho X + Y\rho Y + Z\rho Z\big),
\end{equation}
where \(X, Y, Z\) are the Pauli operators. Thus the state remains unchanged with probability \(1-p\), and with probability \(p\) it is perturbed by a random Pauli error.

This model generalizes to two-qubit gates by uniformly sampling from the 15 non-identity tensor products of Pauli operators. This number arises because the two-qubit Pauli group contains $4 \times 4 = 16$ products of $\{I, X, Y, Z\}$, and excluding the identity $I \otimes I$ leaves exactly 15 non-trivial elements (e.g., $X \otimes I$, $Z \otimes Y$, $Y \otimes X$). The two-qubit depolarizing noise is given by \cite{watrous2018theory}
\begin{equation}
\mathcal{E}_{\mathrm{dep}}^{(2)}(\rho) = (1-p_2)\rho + \frac{p_2}{15}\sum_{i=1}^{15} P_i\,\rho\,P_i^{\dagger},
\end{equation}
where \(\{P_i\}_{i=1}^{15}\) enumerates the non-identity two-qubit Pauli operators. In our framework, depolarizing noise is applied via the Qiskit \emph{NoiseModel} on the two-qubit gates used in the state-preparation, oracle, and diffuser circuits. Following IBM’s example, we apply a depolarizing channel with probability \(p_2=0.02\) to each CNOT gate~\cite{ibm_aer_primitives}.

\subsection{Readout Noise}

Readout noise accounts for classical misassignment errors that occur during the final measurement stage of a quantum circuit. Unlike gate noise, which affects the evolution of quantum states, readout errors arise when the quantum state is projected onto the computational basis and the outcome is misreported. This noise source can be particularly impactful in shallow-depth circuits, such as those used in Grover-style algorithms, where measurement errors may dominate overall performance degradation.

We model measurement errors as a classical assignment matrix noise model applied independently to each qubit, using Qiskit’s \emph{ReadoutError} with the assignment matrix
\begin{equation}
A=
\begin{bmatrix}
P(0|0) & P(0|1) \\
P(1|0) & P(1|1)
\end{bmatrix}
=
\begin{bmatrix}
0.95 & 0.05 \\
0.10 & 0.90
\end{bmatrix}.
\end{equation}
Here \(P(i|j)\) denotes the probability of reporting outcome \(i\) when the true state is \(j\). The entries of \(A\) follow IBM’s noise-modeling example and represent a realistic, mildly pessimistic operating point~\cite{qiskit_noise_models}. This avoids overstating performance, since lower readout error would only improve the reported results. The asymmetry \(P(0|1)>P(1|0)\) is consistent with superconducting-qubit readout, where relaxation during measurement makes \(\lvert 1\rangle\!\to\!\lvert 0\rangle\) misassignment more likely. In our simulations, a uniform readout error is applied to all measured qubits via the Qiskit \emph{NoiseModel}.

\section{Simulation Results}
\label{Results}

This section presents a performance evaluation of the proposed hybrid quantum–classical detector (Hybrid-GAS) in the RIS-assisted SC-FDE system described in Section~\ref{sec:system_model}. The end-to-end channel is assumed to follow a frequency-selective Rayleigh fading model with a uniform power delay profile (PDP). Additive white Gaussian noise (AWGN) is modeled as a zero-mean complex Gaussian random variable with variance $\sigma_n^2 = 1/\mathrm{SNR}$, and perfect channel estimation is assumed.
 Unless otherwise specified, the channel comprises $L=3$ taps, and the RIS phases are configured to align with the central tap. BPSK modulation is adopted throughout.

While the proposed framework applies to larger $N$, our evaluation focuses on moderate problem sizes representative of current NISQ capabilities and tractable for classical quantum-circuit simulation. Increasing $N$ enlarges the $N\times N$ QUBO coefficient matrix $\mathbf{Q}'$ in~\eqref{eq:final_QUBO}, thereby increasing the number of controlled-phase rotations required for phase encoding and, consequently, the circuit depth. Since current quantum-circuit simulators and NISQ hardware support only limited qubit counts and depths, such growth quickly renders large-$N$ instances intractable for near-term quantum execution.

All quantum circuit simulations are performed in IBM Qiskit using the Aer simulator on classical hardware. Both
ideal FTQC circuits and noisy executions are considered. Noise
is introduced via Qiskit’s \emph{NoiseModel}, calibrated to typical IBM Quantum backend
specifications; specifically, depolarizing gate noise and asymmetric readout errors are
included, as detailed in Section~\ref{sec:quantum_noise}.
\normalsize
\setlength{\abovecaptionskip}{0pt}
\setlength{\belowcaptionskip}{0pt}
\begin{figure}
\includegraphics[width=1\linewidth]{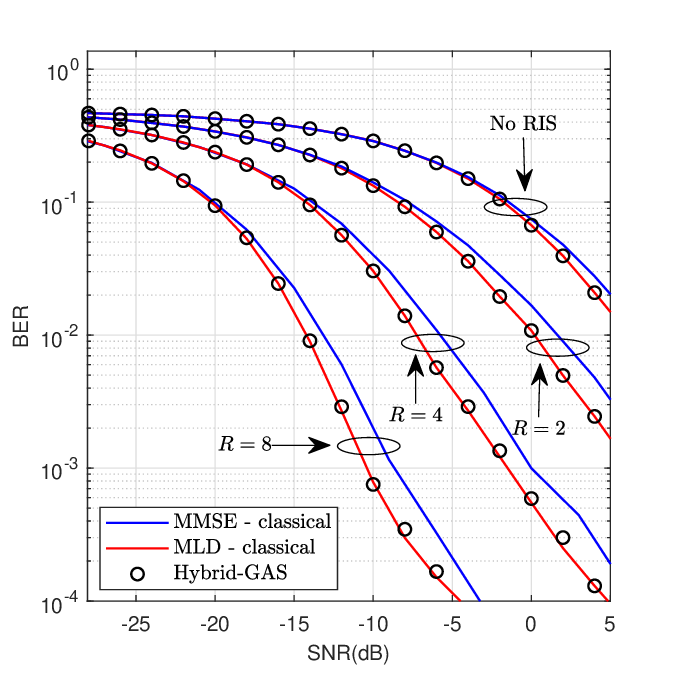}
    \centering
\caption{BER versus SNR for MMSE, MLD, and Hybrid-GAS detectors with and without RIS, shown for $R \in \{2,4,8\}$.}

    \label{fig:ris_gain}
\end{figure}

Fig.~\ref{fig:ris_gain} compares the BER performance of the proposed Hybrid-GAS detector with classical MMSE and MLD baselines for different RIS sizes. Simulations are conducted under ideal conditions with $N=3$. With MMSE-based initialization, Hybrid-GAS consistently outperforms the MMSE baseline and tracks the performance of MLD across all configurations. The results also demonstrate significant gains from increasing the RIS size $R$, enabling coherent signal combining at the receiver and thereby improving both signal quality and overall system performance.
\begin{figure}
    \includegraphics[width=1\linewidth]{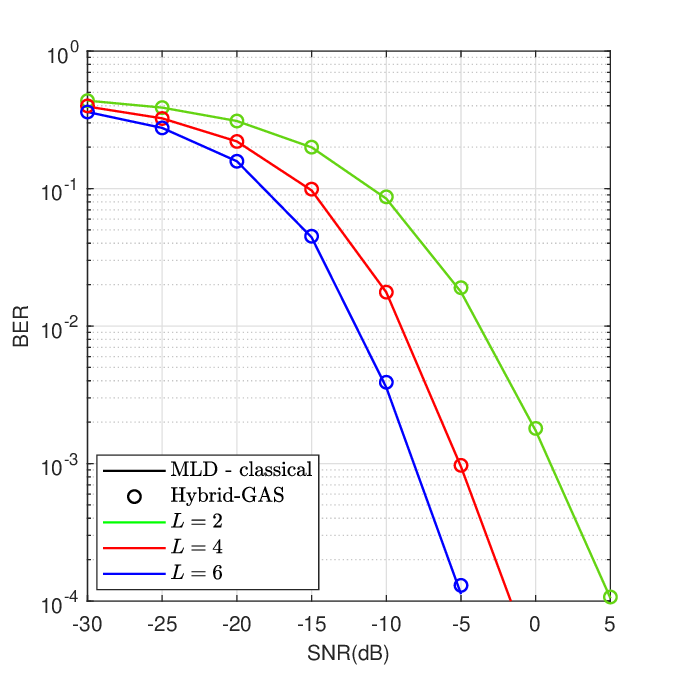}
    \centering
\caption{BER versus SNR for Hybrid-GAS and MLD with $L \in \{2,4,6\}$, $N=6$, and $R=4$.}
    \label{fig:ris_taps}
\end{figure}

In Fig.~\ref{fig:ris_taps}, the BER performance of the Hybrid-GAS detector and the optimal MLD for different numbers of end-to-end channel taps ($L$) is presented. Simulations are performed with $N=6$ and $R=4$ in ideal conditions. The Hybrid--GAS closely tracks the MLD benchmark across all $L$ values. Increasing $L$ enhances performance by providing additional multipath diversity, which yields a steeper BER slope at high SNRs. These results confirm the effectiveness of the proposed approach in highly frequency-selective channels, where leveraging multipath diversity is critical for reliable detection.

\setlength{\abovecaptionskip}{0pt}
\setlength{\belowcaptionskip}{0pt}
\begin{figure}
    \includegraphics[width=1\linewidth]{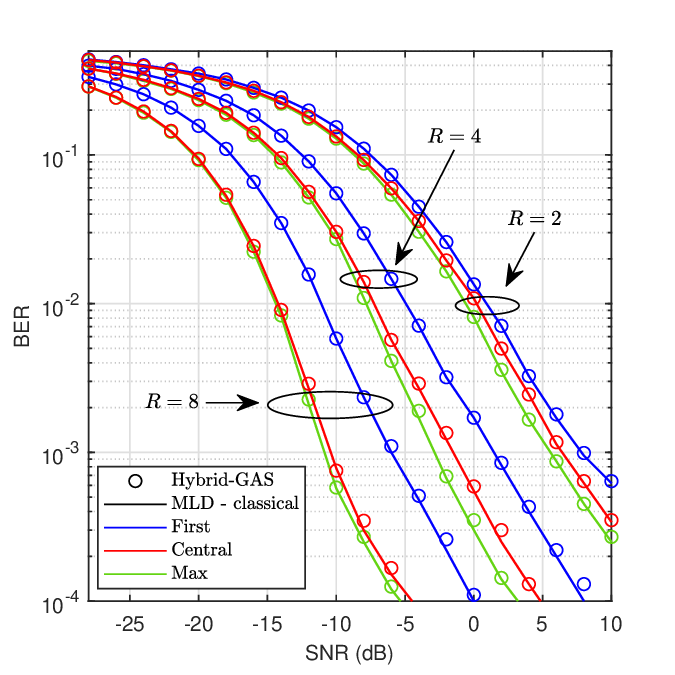}
    \centering
\caption{BER versus SNR for Hybrid-GAS and MLD under RIS tap-compensation configurations of first tap, central tap, and maximum tap.}

    \label{fig:ris_compensation}
\end{figure}

Fig.~\ref{fig:ris_compensation} shows the BER performance of the proposed detector under three RIS tap-compensation strategies discussed in Section~\ref{sec:system_model}, evaluated under ideal conditions. Compensating for the tap with maximum power consistently provides the best performance, while first-tap compensation is least effective under a uniform PDP. As the number of RIS elements increases, central-tap compensation approaches maximum-tap performance due to the convolutional structure of the end-to-end CIR, which concentrates energy around the central tap. Across all strategies, the Hybrid-GAS closely tracks the MLD benchmark with minimal loss, confirming its practical viability in RIS-assisted frequency-selective channels under FTQC-like conditions.

\setlength{\abovecaptionskip}{0pt}
\setlength{\belowcaptionskip}{0pt}
\begin{figure}
\includegraphics[width=1\linewidth]{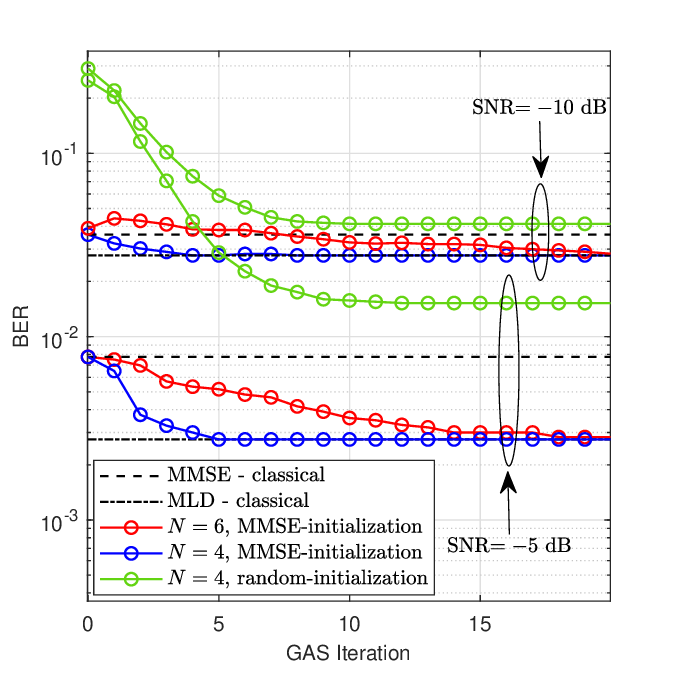}
    \centering
\caption{BER versus GAS iteration at $\mathrm{SNR} \in \{-10,-5\}\,\mathrm{dB}$ for block lengths $N \in \{4,6\}$ and $R=4$, comparing MMSE-based and random (conventional GAS) initializations.}

    \label{fig:convergence}
\end{figure}

The convergence behavior of Hybrid--GAS for block lengths $N=4$ and $N=6$ at $\mathrm{SNR}=-5,\mathrm{dB}$ and $-10,\mathrm{dB}$ is illustrated in Fig.~\ref{fig:convergence}. Simulations are conducted under both random and MMSE threshold initialization, where random initialization 
represents the conventional GAS baseline. The MMSE-based initialization corresponds to iteration index $i=0$, i.e., before any Grover iterations are applied. 
Across both SNR values, $N=4$ converges faster than $N=6$, requiring about five iterations compared to roughly eighteen iterations for $N=6$. This difference reflects the smaller search space for $N=4$, which allows faster convergence.

\setlength{\abovecaptionskip}{0pt}
\setlength{\belowcaptionskip}{0pt}
 \begin{figure}
\includegraphics[width=1\linewidth]{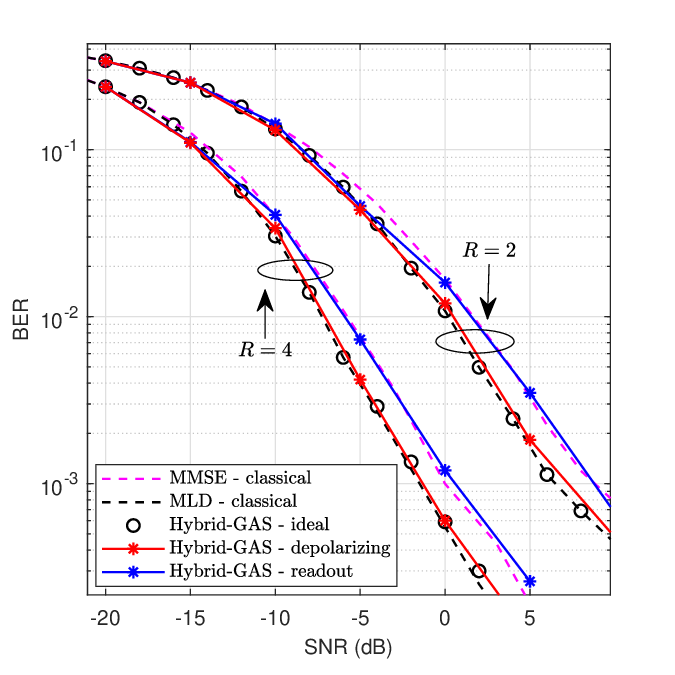}
    \centering
\caption{BER versus SNR of Hybrid-GAS under ideal, depolarizing, and readout noise scenarios, for $N \in \{3,4\}$ at $\mathrm{SNR} \in \{-10,-5\}\,\mathrm{dB}$, with $L=3$, $R=4$, random and MMSE threshold initialization, and RIS central-tap compensation.}

\label{fig:noise_impact}
\end{figure}
 
 Fig.~\ref{fig:noise_impact} evaluates the robustness of the Hybrid-GAS detector under two representative quantum noise models, depolarizing noise and readout noise. Under depolarizing noise, the BER shows only limited degradation. This resilience reflects both the shallow depth of GAS circuits and the MMSE-based threshold initialization, which often accelerates convergence. In addition, depolarizing errors are symmetric and unbiased, so they do not introduce systematic distortions into amplitude amplification. Since the final cost $E(\mathbf{b})$ is recomputed classically after measurement, the detection rule remains stable even under moderate state degradation. In contrast, readout noise causes more pronounced degradation at higher SNRs due to misclassification of low-probability error events. Even so, the BER remains close to that of the initial MMSE threshold, confirming that Hybrid-GAS maintains strong detection capability under realistic quantum measurement errors. The noise parameters, detailed in Section~\ref{sec:quantum_noise}, are taken from IBM’s Qiskit noise examples and are intended to reflect practical values for current superconducting devices, ensuring that the analysis captures realistic NISQ-era limitations.

\section{Conclusion}
\label{conc}
In this work, a hybrid quantum–classical detection framework was developed for RIS-assisted SC-FDE systems, in which the MLD objective is reformulated as a QUBO and solved using GAS.
 A frequency-domain MMSE equalizer was integrated to initialize the adaptive threshold, providing a reliable starting point that accelerates convergence of the quantum search. The quantum resources of Hybrid--GAS were quantified in terms of register widths and gate counts, with scaling expressed explicitly in the block length $N$, channel memory $L$, and the number of qubits $m$ used to represent the shifted cost in two’s complement form. This characterization provides a clear profile of the algorithmic cost for block transmission over frequency-selective channels. Simulations showed that the detector achieves near-optimal performance under ideal conditions, while by construction the approach inherits Grover’s quadratic speedup, reducing query complexity from $\mathcal{O}(M^{N})$ to $\mathcal{O}\left(\sqrt{M^{N}}\right)$. Evaluations across different RIS sizes and channel lengths further confirmed robustness in frequency-selective fading channels and scalability with system dimensions. Under noisy conditions, depolarizing errors had negligible impact, while readout errors caused only modest degradation, with performance remaining close to the MMSE baseline.


\bibliographystyle{IEEEtran}
\bibliography{References}

\end{document}

%% file: Acronyms.tex
\newacronym{ISI}{ISI}{intersymbol interference}
\newacronym{6G}{6G}{sixth-generation}
\newacronym{RIS}{RIS}{reconfigurable intelligent surfaces}
\newacronym{MLD}{MLD}{maximum likelihood detector}
\newacronym{SC-FDE}{SC-FDE}{single-carrier frequency domain equalization}
\newacronym{BS-RIS-UE}{BS-RIS-UE}{base station-ris-user equipment}
\newacronym{SC-FDMA}{SC-FDMA}{single carrier frequency division multiple access}
\newacronym{IM}{IM}{index modulation}
\newacronym{PAPR}{PAPR}{peak-to-average power ratio}
\newacronym{PSK}{PSK}{phase shift keying}
\newacronym{PEP}{PEP}{pairwise error probability}
\newacronym{CPSC-RIS}{CPSC-RIS}{cyclic-prefixed sc-ris}
\newacronym{SNR}{SNR}{signal to noise ratio}
\newacronym{SC-FDMA-IM}{SC-FDMA-IM}{single carrier frequency division multiple access with index modulation}
\newacronym{BER}{BER}{bit error rate}
\newacronym{ISAC}{ISAC}{integrated sensing and communications}
\newacronym{ِAVs}{ِAVs}{autonomous vehicle}
\newacronym{ِV2V}{V2V}{vehicle-to-vehicle}
\newacronym{ِV2I}{V2I}{vehicle-to-infrastructure}
\newacronym{ِV2P}{V2P}{vehicle-to-pedestria}
\newacronym{ML}{ML}{maximum liklihood}
\newacronym{OFDMA}{OFDMA}{orthogonal frequency division multiple access}
\newacronym{LTE}{LTE}{long-term evolution}
\newacronym{5G}{5G}{fifth generation mobile networks}
\newacronym{DFT}{DFT}{discrete Fourier transform}
\newacronym{IDFT}{IDFT}{inverse discrete fourier transform}
\newacronym{MIMO}{MIMO}{multiple-input multiple-output}
\newacronym{SM}{SM}{spatial modulation}
\newacronym{AG}{AG}{antenna grouping}
\newacronym{SIM}{SIM}{subcarrier index modulation}
\newacronym{TIM}{TIM}{time index modulation}
\newacronym{IoT}{IoT}{internet of things}
\newacronym{BS}{BS}{base station}
\newacronym{V2V}{V2V}{vehicle-to-vehicle}
\newacronym{QAM}{QAM}{quadrature amplitude modulation}
\newacronym{IBI}{IBI}{inter-block interference}
\newacronym{CP}{CP}{cyclic prefix}
\newacronym{MMSE}{MMSE}{minimum mean square equalizer}
\newacronym{SISO}{SISO}{single-input single-output}
\newacronym{MISO}{MISO}{multiple-input single-output}
\newacronym{SC}{SC}{single-carrier}
\newacronym{LAS}{LAS}{likelihood ascent search}
\newacronym{SC-IM}{SC-IM}{single-carrier index modulation}
\newacronym{FDE}{FDE}{frequency domain equalization}
\newacronym{OFDM-IM}{OFDM-IM}{orthogonal frequency division multiplexing with index modulation}
\newacronym{IMMA}{IMMA}{index modulation multiple access}
\newacronym{OFDM}{OFDM}{orthogonal frequency division multiplexing}
\newacronym{ABER}{ABER}{average bit error rate}
\newacronym{NOMA-IM}{NOMA-IM}{non-orthogonal multiple access with index modulation}
\newacronym{URLLC}{URLLC}{ultra-reliable low-latency communication}
\newacronym{AV}{AV}{autonomous vehicle}
\newacronym{V2X}{V2X}{vehicle-to-everything}
\newacronym{V2I}{V2I}{vehicle-to-infrastructure}
\newacronym{V2P}{V2P}{vehicle-to-pedestrian}
\newacronym{NLOS}{NLOS}{non-line-of-sight}
\newacronym{QPSK}{QPSK}{quadrature phase shift keying}
\newacronym{FFT}{FFT}{fast fourier transform}
\newacronym{IFFT}{IFFT}{inverse fast fourier transform}
\newacronym{DFE}{DFE}{decision feedback equalizer}
\newacronym{mmWave}{mmWave}{millimeter wave}
\newacronym{CRB}{CRB}{cramér-rao bound}
\newacronym{AR}{AR}{augmented reality}
\newacronym{VR}{VR}{virtual reality}
\newacronym{LoS}{LoS}{line of sight}
\newacronym{spectral efficiency}{SE}{spectral efficiency}
\newacronym{energy efficiency}{EE}{energy efficiency}
\newacronym{vehicular communication}{VC}{vehicular communication}
\newacronym{non-terrestrial networks}{NTN}{non-terrestrial networks}
\newacronym{NOMA}{NOMA}{non-orthogonal multiple access}
\newacronym{D2D}{D2D}{device-to-device communication}
\newacronym{frequency-selective fading}{FSF}{frequency-selective fading}
\newacronym{localization}{LOC}{localization}
\newacronym{tracking}{TRK}{tracking}
\newacronym{multicarrier}{MC}{multicarrier}
\newacronym{DFT-s-OFDM}{DFT-s-OFDM}{discrete fourier transform-spread-ofdm}
\newacronym{Golay}{Golay}{Golay complementary sequences}
\newacronym{Doppler}{Doppler}{doppler shift}
\newacronym{CP-SC}{CP-SC}{cyclic prefixed single carrier}
\newacronym{Impulse Based Single Carrier UWB}{IBSC-UWB}{impulse based single carrier ultra wide band}
\newacronym{Multi-Carrier UWB}{MC-UWB}{multi-carrier ultra wide band}
\newacronym{mutual information}{MI}{mutual information}
\newacronym{NCS}{NCS}{networked control systems}
\newacronym{ZFE}{ZFE}{zero-forcing equalization}
\newacronym{ICI}{ICI}{inter-carrier interference}
\newacronym{IRS}{IRS}{intelligent reflecting surfaces}
\newacronym{LIS}{LIS}{large intelligent surfaces}
\newacronym{SC-TDE}{SC-TDE}{single-carrier time-domain equalization}
\newacronym{RF}{RF}{radio frequency}
\newacronym{CFO}{CFO}{carrier frequency offset}
\newacronym{MM}{MM}{minorization-maximization}
\newacronym{CPSC}{CPSC}{cyclic-prefixed single-carrier}
\newacronym{UE}{UE}{user equipment}
\newacronym{CIR}{CIR}{channel impulse response}
\newacronym{IID}{i.i.d}{independent and identically distributed}
\newacronym{PDP}{PDP}{power delay profile}
\newacronym{OTFS}{OTFS}{orthogonal time frequency space}
\newacronym{SC-OTFS}{SC-OTFS}{single carrier orthogonal time frequency space}
\newacronym{BPSK}{BPSK}{binary phase shift keying}
\newacronym{QoS}{QoS}{quality of service}
\newacronym{AWGN}{AWGN}{additive white gaussian noise}

\newacronym{FDMA}{FDMA}{frequency division multiple access}
\newacronym{MA}{MA}{multiple access}
\newacronym{WB}{WB}{wide band}
\newacronym{FRC}{FRC}{frequency correlation}
\newacronym{TDE}{TDE}{time domain equalization}
\newacronym{PSD}{PSD}{power spectral density}
\newacronym{PDF}{PDF}{probability density function}
\newacronym{UWB}{UWB}{ultra wide band}
\newacronym{MC}{MC}{multi-carrier}
\newacronym{MEMS}{MEMS}{micro-electro-mechanical systems}
\newacronym{PIN}{PIN}{positive-intrinsic-negative diode}
\newacronym{FPGA}{FPGA}{field-programmable gate array}
\newacronym{PCB}{PCB}{printed circuit board}
\newacronym{ZP}{ZP}{zero padding}
\newacronym{CDF}{CDF}{cumulative distribution function}
\newacronym{CLT}{CLT}{central limit theorem}

\newacronym{CSI}{CSI}{channel state information}
\newacronym{XR}{XR}{extended reality}

\newacronym{AI}{AI}{artificial intelligence}